# The Effect of Baroque Music on the PassPoints Graphical Password

Haichang Gao, Zhongjie Ren, Xiuling Chang, Xiyang Liu
Software Engineering Institute
Xidian University
Xi'an, Shaanxi 710071, P.R.China
hchgao@xidian.edu.cn

Uwe Aickelin
School of Computer Science
The University of Nottingham
Nottingham, NG8 1BB, U.K.
uxa@cs.nott.ac.uk

## ABSTRACT

Graphical passwords have been demonstrated to be the possible alternatives to traditional alphanumeric passwords. However, they still tend to follow predictable patterns that are easier to attack. The crux of the problem is users' memory limitations. Users are the weakest link in password authentication mechanism. It shows that baroque music has positive effects on human memorizing and learning. We introduce baroque music to the PassPoints graphical password scheme and conduct a laboratory study in this paper. Results shown that there is no statistic difference between the music group and the control group without music in short-term recall experiments, both had high recall success rates. But in long-term recall, the music group performed significantly better. We also found that the music group tended to set significantly more complicated passwords, which are usually more resistant to dictionary and other guess attacks. But compared with the control group, the music group took more time to log in both in short-term and long-term tests. Besides, it appears that background music does not work in terms of hotspots.

## Categories and Subject Descriptors

H.1.2 [**User/Machine Systems**]: Human factors; D.4.6 [**Security and Protection**]: Authentication

## General Terms

Security, Human Factors

## Keywords

Graphical password, Baroque music, Memorability, PassPoints

## 1. INTRODUCTION

Alphanumeric passwords are widely used in identity authentication to protect users' privacy. But the *password problem* arises because such passwords are expected to meet two conflicting requirements: (1) Passwords should be easy to remember, and the user authentication protocol should be executable quickly and easily. (2) Passwords should be secure, i.e. they should be random-looking and should be hard to guess; they should be changed frequently, and should be different for multi-accounts; they should not be written down or stored in plain text. Meeting these conditions is almost impossible for humans, with the result that the use of alphanumeric passwords was putted in dilemma: that long complicated passwords are hard for people to remember, while shorter ones are susceptible to attack.

Graphical passwords have been proposed as an alternative to textual passwords with their advantages in usability and security. The main motivation is that the psychologists have shown that in both recognition and recall tasks, images are more memorable than words or sentences [15, 17]. It is conceivable that humans would be able to remember stronger passwords of a graphical nature. However, users still tend to choose passwords that are memorable in some way, which means that the graphical passwords still tend to follow predictable patterns that are easier for attackers to exploit [4, 16, 23].

There have been three dominant techniques available of graphical passwords which can be defined as: Drawmetrics, Locimetrics and Cognometrics [5, 22]. PassPoints is a representative Locimetric scheme of particular interest and worthy of extensive study. In PassPoints, passwords consist of a sequence of several click-points on a given image, and *hotspots* is a primary security problem [2, 7].

Literatures reveal that users are the 'weakest link' in password authentication, probably due to their memory limitations [18, 20, 29]. Psychological and physiological studies indicate that baroque music has positive effects of great importance on human memorizing and learning [8, 10]. In this paper, we investigate the novel idea of introducing background baroque music to the PassPoints graphical password scheme with the purpose of alleviating users' memory burden and improving usable security. A laboratory study was conducted to explore the efficiency of background baroque music on memorizing graphical passwords. We are also interested in whether the background music has other effects on graphical password, like the login time and the password complexity.

The results of our empirical study are very encouraging in PassPoints scheme. The music group coped significantly better than the group without music when recalling pass-words after one week. The music group also tended to set significantly more complicated passwords. This appeared to suggest that the applied music could improve memorability of PassPoints password. Besides, the background music had no significant influence on login times.

The remainder of the paper if outlined as follows: Section 2 reviews graphical password schemes and Baroque music. Sections 3 and 4 describe the methodology of our studies and present the results respectively. Section 5 discusses the experimental results. Conclusion and future work are addressed in Section 6.

## 2. RELATED WORKS

### 2.1 Graphical Passwords

The ubiquity of graphical interfaces for applications and input devices, such as the mouse, stylus and touch-screen, has enabled the emergence of graphical authentications. There have been three kinds of dominant techniques available which can be defined as: Drawmetrics (DAS [14], BDAS [9], YAGP [11]), Locimetrics (Blonder [1], PassPoints [27]) and Cognometrics (Deja Vu [6], Passfaces [9], ColorLogin [12]) [5, 22].

Drawmetrics systems require users to reproduce a pre-drawn outline drawing on a grid. DAS is a typical drawmetric scheme based purely on recall, and requires the user to create a unique image on a drawing grid [14]. BDAS is a variant of DAS [9], which can encourage users to set strong passwords and enhance memorability by introducing background images. YAGP proposed a modification to DAS where approximately correct drawings can be accepted, based on Levenshtein distance string matching and trend quadrants of pen strokes [11]. As consequences of this approximation algorithm, a finer grid may be used.

Originating in Blonder's work, the Locimetrics approach involves users choosing several sequential locations in an image [1, 13]. PassPoints [27] is a representative scheme of this category, where users may choose any place in the image as a password click point. This scheme was found that although relatively usable, security analyses find it vulnerable to hotspots and simple patterns within images [2, 23]. To reduce the security impact of hotspots, CCP (Cued Click-Points) [3] and PCCP (Persuasive-CCP) [2] are proposed.

In the Cognometrics systems, users must recognize the target images embedded amongst a set of distractor images. This category includes Deja Vu [6] based on abstract images, Passfaces which relies on face recognition [19] and ColorLogin [12] using multiple background color to decrease login time. Memorability for abstract images in Deja Vu was found to be only half as good as that for photographic images with a clear central subject [26]. User studies by Valentine have shown that Passfaces has a high degree of memorability [24, 25], but Davis found that people tended to select faces of their favorite [4]. ColorLogin uses background color to decrease login time. Multiple colors are used to confuse the peepers, while not burdening the legitimate users [12]. Meanwhile, the scheme is resistant to shoulder surfing and intersection attack to a certain extent. However, the hotspot is still a problem that needs addressing.

It can be concluded that most graphical passwords either tend to follow predictable patterns or have a low degree of memorability. The crux of the problem is the users' memory limitations. Extensive researches have shown that Baroque music has different uses for education and therapy [27]. Our particular interest is to explore the role of music in learning and memorizing graphical passwords.

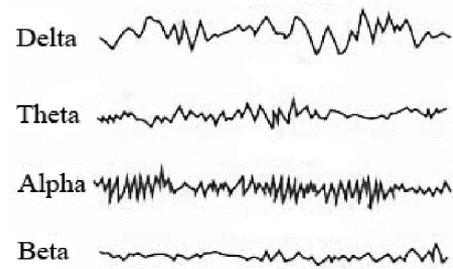

**Figure 1: Different brain waves.**

### 2.2 The Baroque Music

As human memory capacity is unlikely to increase significantly over the next few years, creating a nice environment for memorizing passwords might alleviate users' burden. Our investigation was mainly motivated by scientific literature in psychology and physiology. There are demonstrations that music can improve memory and in what flows we will illustrate it. Georgi Lozanov made remarkable impact in integrating music into teaching practice [8, 21]. He created a teaching method called *Suggestopedia*, wherein the use of background music, particularly the baroque music with a rate of 50 to 70 beats per minute (BPM), is a cornerstone of accelerated learning techniques. We will briefly review the researches into the effects of music on learning in this subsection.

There are various researches with regards to our brain and different cycles it works in. As shown in Figure 1, the brain wave can be divided into four types according to the frequency. They are named beta, alpha, theta and delta in decreasing order, which represent different states of mind. The frequency of beta state is 13-25 CPS (Cycles per Second), alpha 8-12 CPS, theta 4-7 CPS and delta 0.5-3 CPS. When we are wide awake and alert, figuring out complex problems and talking, our brain will probably stay at beta state, which characterizes logical thought, analysis and action. This is the brain wave of our conscious mind and thus not the best state for stimulating our long-term memory [8]. Instead, alpha lets us reach our subconscious in which most information we learn will be stored. It is a state of relaxed alertness, facilitating inspiration, fast assimilation of facts and heightened memory. While alpha characterizes relaxation and meditation, the theta deep meditation and reverie. The theta state can be best described as being the *twilight zone between being fully awake and fully asleep* and the delta state is reached when we are in deep sleep [8]. In conclusion, the alpha state is optimal for learning and memorizing.

Baroque music can help the brain produce alpha waves, and information imbued with music has a greater likelihood of being encoded in the long-term memory by the brain. That is why accelerated learning techniques introduce music into the learning process. For example, *Mozart Effect* [21] is a phenomenon that music has a positive effect on learning and memory.

In the previous subsection, we have found that human memory limitations have caused lots of security problems. We bring background baroque music to the PassPoints graphical password scheme and do an investigation to check whether it can improve users' memory or induce users to set stronger passwords.

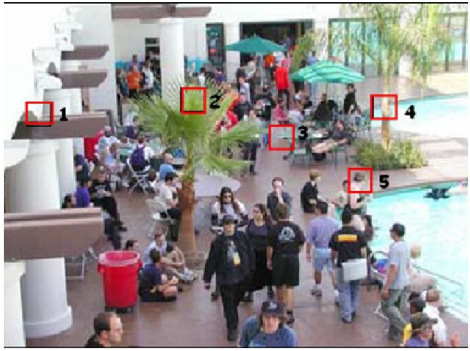

**Figure 2: Passwords in PassPoints with length being 5.**

## 3. EXPERIMENTS

For the purpose of collecting and analyzing the success rate, user habits, and login time automatically, we reproduce the scheme which is intentionally very closely modeled after the original PassPoints [27]. We still adopt the name PassPoints for convenience.

Users are required to select several positions in a single image as their passwords and click close to the chosen points in correct order and within a tolerance distance for authentication. To maintain compatibility with previous studies [27] as much as possible, PassPoints application used pool images (315 × 236 pixel) and tolerance area of 20 × 20 pixels [28]. For example, the password in Figure 2 contains five click points orderly labeled by small red rectangle.

A population of 28 participants was invited to the experiment study. All the participants were university students and the average age of the participants was 26 years old. We hypothesized that background music could improve humans memory and then induced people to choose more complex passwords and take less time to log in. This study used a between-subjects design, 14 participants were assigned to the control group without background music and the other half to the music group. None of them had previously used PassPoints password. We chose the baroque music suggested by Lozanov with a rate of 50 to 70 BMP as the background music. The speaker volume was set to 30 to 40 decibels as suggested.

There were two lab-based sessions in our user study. Session 1 was a short-term one, taking about two hours. At the beginning of Session 1, each participant was asked to read an instruction document. This provided information of their activities on the experiments and helped them know how PassPoints works. To make the rules clearer, an example was included. Then participants were required to complete the registration and login of PassPoints. People were asked to reenter the password to confirm it. After 10 minutes short delay, participants were asked to log in within at most three attempts. In the end, participants need answer a demographic questionnaire collecting information including age, sex and experience on graphical passwords. One week later, at Session 2, all the participants returned to the lab and tried to log in the scheme within three attempts using their previously created passwords.

**Table 1: The login times (Seconds) for both groups in each session of the study**

| Group | | Avg. | t-test | S.d. | Max | Min |
|---|---|---|---|---|---|---|
| 10 Minutes test | No music | 16.5 | No | 8.6 | 33 | 7 |
| | With music | 15.3 | | 5.9 | 31 | 10 |
| One week test | No music | 40.6 | - | 22.3 | 87 | 12 |
| | With music | 62.3 | | 49.2 | 213 | 17 |

## 4. RESULTS

Two types of statistical tests were used to evaluate whether differences in the data reflect actual disparity between conditions or whether these may have occurred by chance. A two tails t-test was used for comparing the means of two groups and Fisher's exact test was used to compare recall success rates. In all cases, we regard a value of $P<0.20$ as indicating that the groups being tested are different from each other with at least 80% probability, making the result statistically significant. "No" means not significant in the tables. It indicates that the test revealed no statistically significant difference between the two conditions.

### 4.1 Login Time

Since people were not familiar with graphical passwords and then it usually takes much more time to either create or enter graphical passwords, we are interested in finding a method to reduce it. Previous sections have claimed that background music could improve humans' memory, so we assumed that less time were needed to recall the passwords for the music group. In Table 1, the login time represented the total time spent during the authentication, which began when the login screen first appeared and continued until the user entered their username and password. 'No' means not significant in t-test.

Compared with the control group, the music group took more time to log in a scheme both in short-term and longterm tests. The results of t-tests (two tails) showed that none of the differences were statistically significant. It was worthwhile to note that one person of the music group took 213s to log in PassPoints in the long-term test and data from two groups did not satisfy the requirements of "homogeneity of variances". A t-test was thus not available to test the difference. Excluding the data 213s, the max time was 80s for the music group and the average 48.6s.

### 4.2 Success Rates

We examine success rates as a measure of participants' performance. Table 2 compares the successful recalls in each group. During the recall after ten minutes, the success rates were high on the whole, both 100% success rate indicating that participants' memory was not strongly taxed during this phase.

After one week, the performances of two groups varied in schemes. We found a significant difference between two groups. The music group was significantly more likely to successfully recall the passwords than the control group. In addition, the success rate of the control group decreased from 100% in the previous phase to 35.7% while the success rate of the music group only decreased by 7.1%. P=0.004 in fisher-test means that the groups being tested are different from each other with great probability. It aligns with psychology research which continues to show that certain music advance the long-term memory. The results suggest that the background music could significantly help people remember passwords in long-term memory.

**Table 2: Success rates in each group for PassPoints**

| Group | 10-minute test | | 1-week test | |
|---|---|---|---|---|
| | ratio | Fisher-test | ratio | Fisher-test |
| No music | 100% | P=1 | 37.5% | P=0.004 |
| With music | 100% | | 92.9% | |

**Table 3: Complexity of PassPoints**

| Group | Password length | | | | |
|---|---|---|---|---|---|
| | Avg. | t-test | S.d. | Max | Min |
| No music | 3.79 | t=1.61, P<0.20 | 1.20 | 5 | 1 |
| With music | 4.5 | | 1.05 | 6 | 3 |

## 5. DISCUSSION

Based on the previous results, we now revisit our hypotheses that background music could improve humans' memory and then induced people to choose more complex passwords. This hypothesis was supported in PassPoints considering the password complexity. People in the music condition not only chose significantly more complicated passwords, but also had significantly higher recall success rates in the long-term test. In respect of login time and other aspects, there were some differences but not statistically significant between two groups with and without music.

### 5.1 Recall Errors

People committed different types of error as shown in Table 4. There are three types of error in PassPoints: pwd-Len error, i.e., people forgetting the password length; position error, i.e., people can recall the password length but click outside the tolerance region; and order error, i.e., people can recall the password length and position but mixing up the click-points order. In PassPoints scheme, the nature of many recall failure was down to either forgetting the password length or clicking points outside the tolerance region. In recall errors and especially in position errors, music group had a great advantage over non-music group, probably due to its higher success recall rate in the long-term recall test.

### 5.2 Hotspots

**Table 4: Recall errors in PassPoints**

| Group | | Pwd-len | Position | Order |
|---|---|---|---|---|
| 10 minutes test | No music | 14 | 16 | 2 |
| | With music | 9 | 9 | 3 |
| One week test | No music | 2 | 2 | 1 |
| | With music | 1 | 2 | 0 |

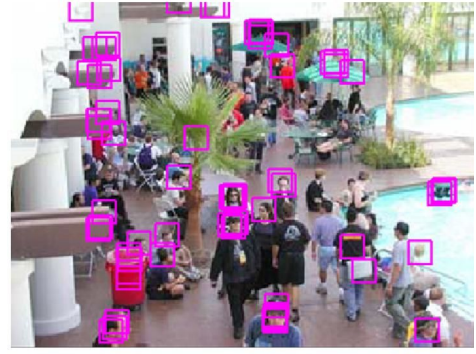

**Figure 3: Hotspots in music group.**

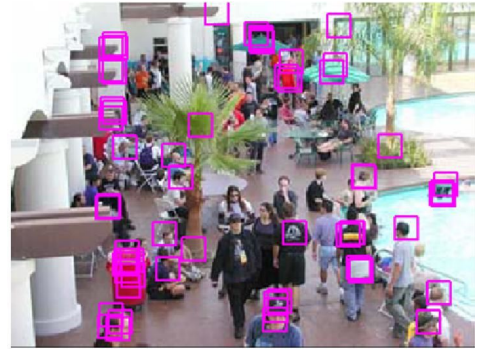

**Figure 4: Hotspots in control group without music.**

Hotspots are areas of the image with higher probability of being chosen by users as individual click-points. Hotspot is a serious security problem in click-based schemes. Figure 3 shows the hotspots distribution of PassPoints passwords in music group, Figure 4 shows the hotspots in control group without music, and Figure 5 shows the total hotspots in two groups. We can see similar slight clustering of click-points in both groups. It appears that background music does not work in terms of hotspots. Points with high visual salience were still more likely to be selected as passwords.

### 5.3 Interpretation of Login Time

Contrary to our expectation, participants in the music condition took a little more time to log in a scheme in two tests. We now take a closer look at this issue to understand where the offset arose. We found that participants in the music group had a slightly higher average username length (5.93 vs. 4.21). Furthermore, username was greater than password in average length, which means that the time to enter the username can not be ignored. Besides, we found that the music group took less attempts to log in the short-recall test (1.21 vs. 1.38) and partially hold in the long-term test.

Therefore, it might be the time to recall and enter the username that results in the slight difference between two groups in login time. To evaluate the login time more precise, another program collecting the time to recall and enter passwords is necessary.

## 6. CONCLUSIONS

Study results have shown that it is an effective means to introduce baroque music to the PassPoints graphical password scheme considering the password memorability and complexity. With music stimulus, people not only tended to construct significantly more complicated passwords than their counterparts without music, but also performed significantly better in terms of recall success in the long-term tests. This result indicated that the background music improved the memorability of passwords in PassPoints. But in respect of login time and hotspots, there were no statistically significant differences between two groups.

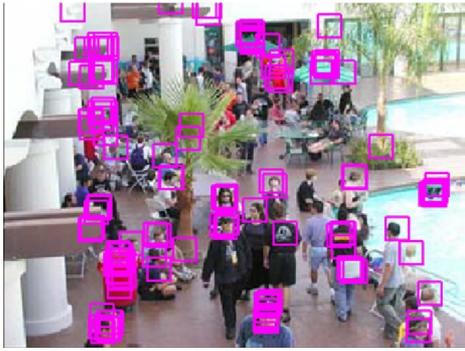

**Figure 5: Total hotspots in two groups.**

We made our study follow the established methods of experimental psychology as much as possible and admitted that it did not reflect the true situation strictly. First, the participants in our study (all of them were university students and very young) only represented a small part of the whole. It was important to get a wider selection of people with various backgrounds in the further studies. Second, the participants had no incentive to perform as if protecting or accessing anything of real-life value to them, therefore it was not difficult to understand that many passwords created in both conditions were weak. Third, the effect of the background music volume remains to be discussed when it was embedded into a scheme. Despite these limitations, our controlled laboratory experiment laid a good foundation to further deep studies.

This work provides a significant extension to the study of security and usability of the click-based PassPoints graphical password. The future work includes a larger scale of studies with careful experimental design and comprehensive study of the Baroque music effect on graphical password.

## 7. ACKNOWLEDGMENTS

The authors would like to thank the reviewers for their helpful and constructive comments of this paper. Project 60903198 supported by National Natural Science Foundation of China.